\documentclass{appolb}
\usepackage{graphicx}
\usepackage{amsmath,amssymb}

\usepackage[backend=bibtex, sorting=none, defernumbers]{biblatex}
\usepackage{hyperref}
\hypersetup{
    colorlinks = true,
    linkcolor = {blue},
    citecolor = {blue},
}
\begin{document}
\title{Net-baryon number fluctuations%
\headtitle{Net-baryon number fluctuations}
\headauthor{C. Schmidt et al.}
\thanks{Presented at the Workshop ``Criticality in QCD and the Hadron Resonance Gas'', 29-31 July 2020, Online.}%
}
\author{C. Schmidt, J. Goswami, G. Nicotra, F. Ziesché
\vspace*{-3mm}\address{Universität Bielefeld, Fakultät für Physik, D-33615 Bielefeld, Germnay}
\\[5mm]
{P. Dimopoulos, F. Di Renzo, S. Singh, K. Zambello
}
\vspace*{-3mm}\address{Dipartimento di Scienze Matematiche, Fisiche e Informatiche, Università di Parma and INFN, Gruppo Collegato di Parma
I-43100 Parma, Italy}
}
\maketitle
\begin{abstract}
The appearance of large, none-Gaussian cumulants of the baryon number distribution is 
commonly discussed as a signal
for the QCD critical point. We review the status of the Taylor expansion of
cumulant ratios of baryon number fluctuations along the freeze-out line and also
compare QCD results with the corresponding proton number fluctuations as measured
by the STAR Collaboration at RHIC. To further constrain the location of a possible QCD
critical point we discuss poles of the baryon number fluctuations in the complex
plane. Here we use not only the Taylor coefficients obtained at zero chemical
potential but perform also calculations of Taylor expansion coefficients of the
pressure at purely imaginary chemical potentials. 
\end{abstract}
  
\section{Introduction}
The phase diagram of Quantum Chromodynamics (QCD) is currently investigated with 
large efforts by means of heavy ion experiments at LHC and RHIC, as well as by 
numerical calculations of lattice regularized QCD. While lattice calculations at
vanishing chemical potential made great progress in the last decades, they are still
harmed by the infamous sign problem at nonzero chemical potential. The two main methods
that are currently used to infer on the QCD phase diagram at nonzero baryon chemical
potential $(\mu_B)$ are indirect, they rely on Taylor expansions of observables at
$\mu_B=0$, or analytical continuations from simulations at imaginary chemical potential 
($\mu_B=i\mu_I$). Methods that allow for a direct sampling of the oscillatory
path integral at $(\mu_B)>0$ are currently investigated, see e.g. \cite{Attanasio:2020spv,
Alexandru:2020wrj}.

\begin{figure}[htb]
\begin{center}
    \includegraphics[width=0.37\textwidth]{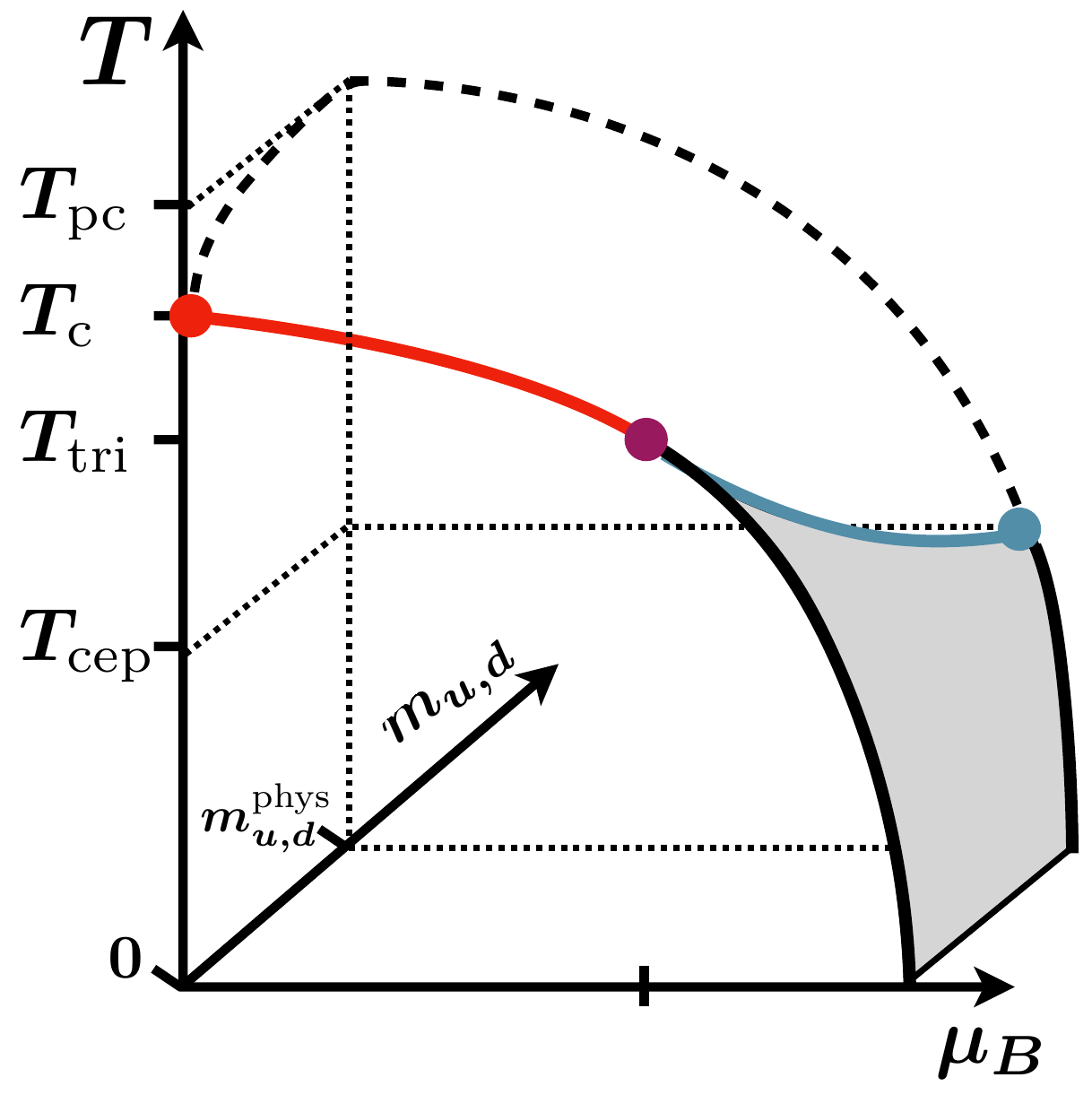}\hfill
    \includegraphics[width=0.58\textwidth]{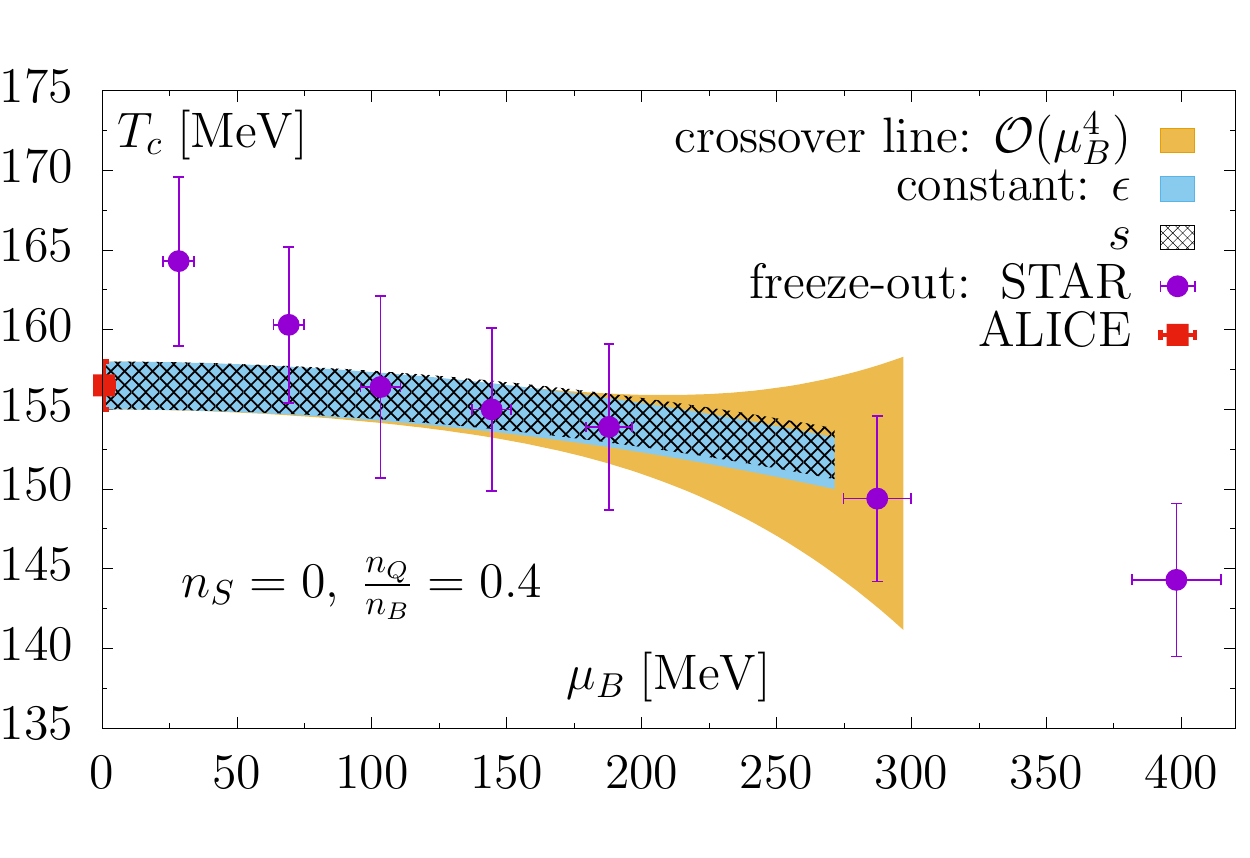}
\end{center}
\caption{Left: Schematic picture of the QCD phase diagram. The chiral limit ($m_{u,d}=0$) 
is shown in the front, whereas the physical mass case is shown in the back. Right:
The chiral crossover line in 2 + 1 flavor QCD, calculated with the constraints
$n_S = 0$ and  $n_Q = 0.4n_B$. It is compared with the line of constant energy density
$\epsilon = 0.42(6) GeV/fm^3$ and the line of constant entropy density $s = 3.7(5)
fm^{-3}$ \cite{Bazavov:2017dus} in the $T-\mu_B$ plane. Also, shown are the chemical 
freeze-out parameters extracted from grand canonical ensemble based fits to hadron yields 
for the ALICE \cite{Andronic:2017pug} and the STAR \cite{Adamczyk:2017iwn} experiments.}
\label{fig:pdiag}
\end{figure}

The two principles that are guiding our understanding of the QCD phase diagram are 
spontaneous chiral symmetry breaking and -- linked to it -- the phenomena of 
quark confinement. Our knowledge on the (2+1)-flavor QCD phase diagram 
based on recent lattice results is summarized in Fig.~\ref{fig:pdiag} (left). 
The variables assigned to the three axes are temperature ($T$), the baryon chemical 
potential ($\mu_B$) and light quark mass ($m_{u,d}$). At low $T$ and low $\mu_B$ the 
chiral symmetry is spontaneously broken and quarks are confined into hadrons. 
Correspondingly, chiral symmetry is restored at high $T$ and high $\mu_B$, where
quarks can move freely.\footnote{For simplicity we are neglecting here various 
superconducting phases at low $T$ and high $\mu_B$, which will not be discussed here.}
Solid red and cyan lines 
indicate a continous phase transition in the universality class of the 3d-O(4) 
symmetric spin model, or the Z(2) symmetric Ising model, respectively. Black lines 
and gray surfaces indicate a discontinuous first order transition. On the temperature axis we also
indicate the pseudo critical transition temperature at physical masses ($T_{pc}$), the
critical temperature in the chiral limit ($T_{c}$), the temperature of the tri-critical
point in the chiral limit ($T_{tri}$) and the temperature of the critical (end-)point at
physical quark masses ($T_{cep}$). It emerges a hierarchy as $T_{pc}>T_{c}>T_{tri}>T_{cep}$. 
The first two temperatures are determined by lattice calculations as $T_{pc}=156(\pm1.5)$ 
MeV \cite{Bazavov:2018mes} and $T_c=132^{+3}_{-6}$ MeV \cite{Ding:2019prx}. 
The variation of $T_{pc}$ with $\mu_B$, 
as indicated by a dashed line, has also been calculated by lattice QCD. We obtain 
\begin{equation}
\frac{T_{pc}(\mu_B)}{T_{pc}(0)}
=1-\kappa_2\left(\frac{\mu_B}{T}\right)^2+\mathcal{O}\left(\frac{\mu_B}{T}\right)^4,
\label{eq:critline}
\end{equation}
where $\kappa_2^B = 0.012(4)$ with a $\mathcal{O}(\mu_B^4)$ correction that vanishes within 
errors \cite{Bazavov:2018mes}. Similar results have been obtained recently in Ref.~\cite{Borsanyi:2020fev}.

In Fig.~\ref{fig:pdiag} (right) we compare the pseudo-critical line with freeze-out temperatures and
chemical potentials obtained from hadron yields measured by STAR \cite{Adamczyk:2017iwn} and
ALICE \cite{Andronic:2017pug}. The hadron yields have been fitted (after feed-down corrections)
to the hadron resonance gas (HRG) model. In its simplest none-interacting version, this model
is based on the mass spectrum of all stable particles and resonances listed in the particle
data booklet, which are taken as an ideal gas in thermal equilibrium at a common temperature
$T_f$, chemical potential $\mu_f$, and volume $V_f$. As these parameters refer to the time in 
the expansion of the fireball from when on its chemical composition does not change anymore, 
they are called chemical freeze-out parameters. We see from Fig.~\ref{fig:pdiag} (right), that the
freeze-out parameters agree well with the chiral crossover line obtained from lattice
QCD. We note, that in order to meet conditions that are found in heavy ion collisions we
have determined our values for the electric $\mu_Q\equiv\mu_Q(\mu_B)$ and strangeness
chemical potentials $\mu_S\equiv\mu_S(\mu_B)$, such that the following conditions for
the net-numbers of conserved charges in the system, $\left<n_Q/n_B\right>=0.4$ and
$\left<n_S\right>=0$, are fulfilled. However, the freeze-out parameters are still model
based. Hence, in the following we want to follow a procedure proposed in 
\cite{Bazavov:2012vg}, that allow for the determination of the freeze-out parameters by a 
direct comparison of lattice QCD to experiment.

\section{Cumulants of net-baryon number}
Higher order cumulants of the net-baryon number are obtained as derivatives of 
the logarithm of the QCD partition functions with respect to the dimension less parameter
$\hat{\mu}_B=\mu_B/T$,
\begin{equation}
    \chi_n^B(T,\mu_B, \mu_Q, \mu_S)
    =\frac{1}{VT^3}
    \frac{\partial^n \ln Z(T,\mu_B,\mu_Q,\mu_S)}{\partial \hat{\mu}_B^n},
\end{equation}
where $\mu_Q$ an $\mu_S$ are the electric charge and strangeness chemical potentials. 
In the same way, we can can also calculate derivatives with respect to $\mu_Q$ and
$\mu_S$, which we denote as $\chi_n^Q$ and $\chi_n^S$, respectively. 

\begin{figure}[htb]
\includegraphics[width=0.5\textwidth]{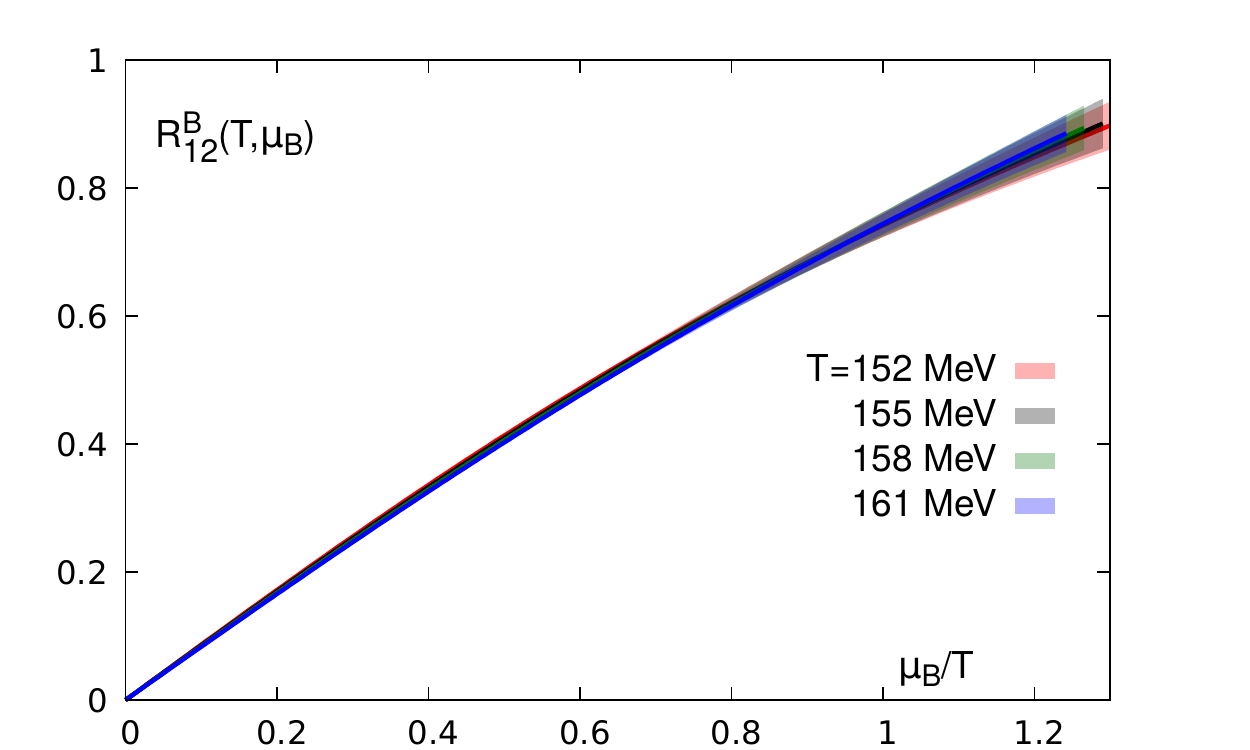}\hfill
\includegraphics[width=0.5\textwidth]{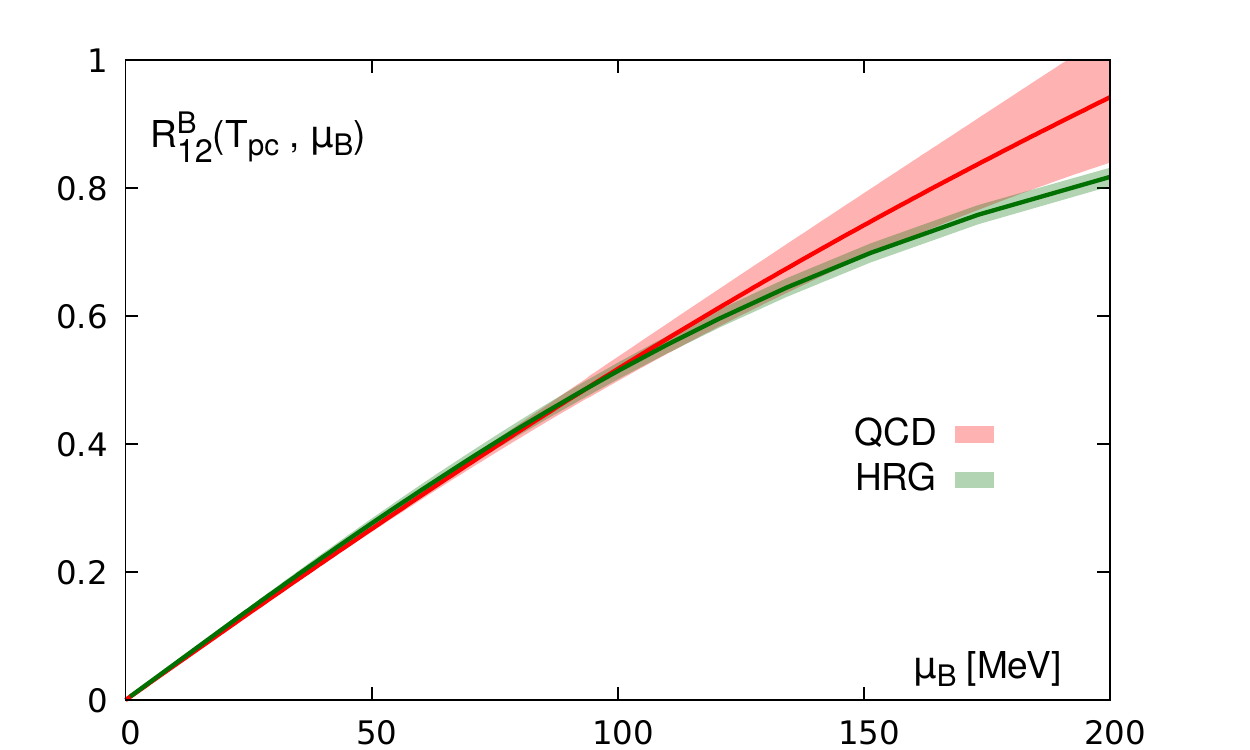}
\caption{Left: Continuum estimate of the cumulant ratio $R_{12}^B$ as function of the chemical
potential, for different temperatures. Right: The same ratio along the pseudo critical line. 
Shown are the QCD and HRG model results.\label{fig:R12B}}
\label{Fig:F2H}
\end{figure}

Aiming on the comparison with the experimental results, we further introduce ratios of 
cumulants of baryon number fluctuations as 
\begin{equation}
    R_{nm}^B=\frac{\chi_n^B}{\chi_m^B}.
\end{equation}
 By using these ratios, the leading order dependence on the freeze-out
volume ($V^f$) is removed. However, among other things fluctuations of the experimentally
observed freeze-out volume might still hinder a comparison to lattice QCD. The first ratio 
we discuss is $R_{12}^B$, which is shown in Fig.~\ref{fig:R12B}
and can be interpreted as the mean of the net-baryon number, normalized by the variance of the
baryon number fluctuations. The presented HotQCD results \cite{Bazavov:2020bjn} are obtained from 
high statistics lattice QCD calculations on $32^3\times 8$ and $48^3\times 12$ lattices, with
(2+1)-flavor of highly improved staggered quarks (HISQ) at physical light and strange quark masses. 
The values in the range $0<\hat\mu_B\lesssim 1.2$ stem from a Taylor expansion of the logarithm of the
partition function about $\hat\mu_B=0$ to $8^{th}$ order in $\hat\mu_B$.
As it is evident from the continuum estimate
shown in Fig.~\ref{fig:R12B} (left), the leading order of $R_{12}^B$ is linear in $\mu_B$. 
We further notice that the ratio is rather independent under the variation of temperature. 
Therefore the ratio has been termed a baryometer \cite{Bazavov:2012vg}. 

The same ratio is shown in Fig.~\ref{fig:R12B} (right), now plotted
along the pseudo critical line as defined in Eq.~(\ref{eq:critline}). Here we compare the QCD result
with the corresponding calculation of a Hadron Resonance Gas (HRG). We see that the HRG model deviates
from QCD only for $\mu_B\gtrsim 150$~MeV. We thus note that for small $\mu_B$ the HRG can be used to
analyse the differences between net-baryon number and net-proton number fluctuations. The latter is the
quantity which is directly accessible by heavy ion experiments. On the other hand, this also means that
we do not see any indication of a diverging baryon number fluctuation ($\chi_2^B$) in the range where we
trust our Taylor expansion, which we would expect in QCD close to a critical point. In this case the
ratio $R_{12}^B$ would decrease and approach zero at the critical point.

\begin{figure}[tb]
\centerline{%
\includegraphics[width=0.55\textwidth]{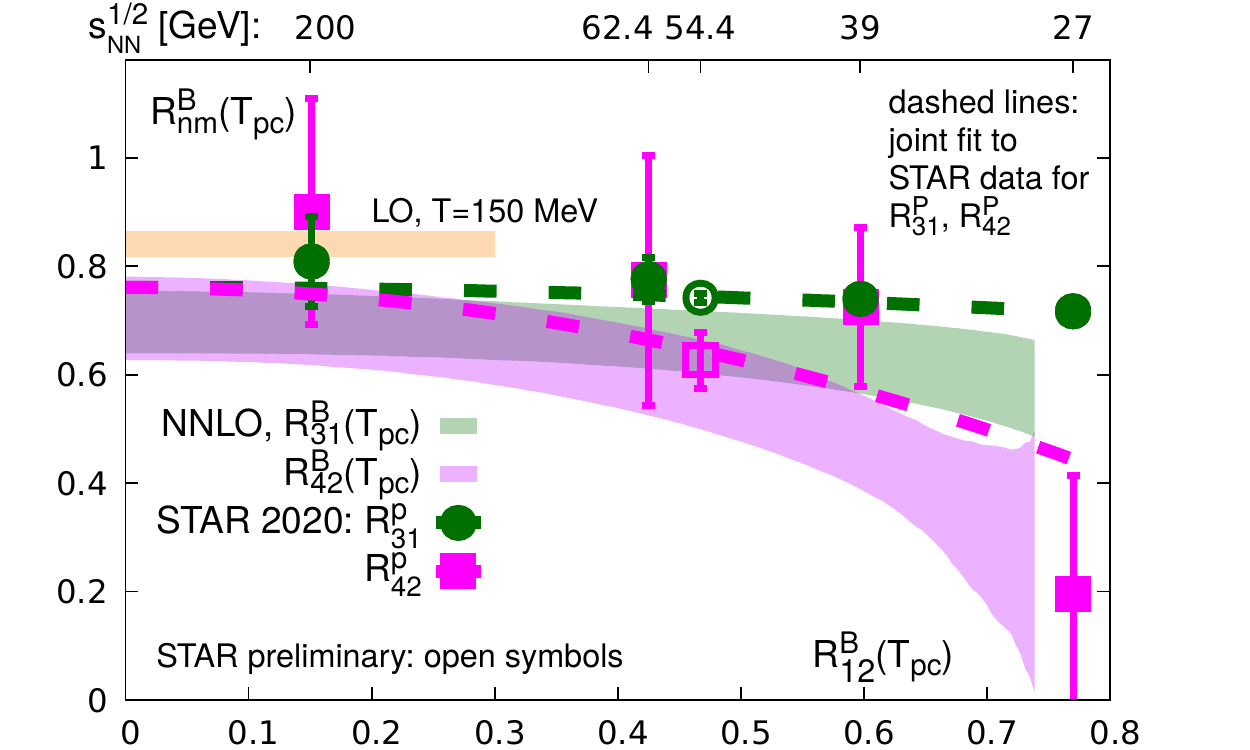}}
\caption{
The cumulant ratios (bands) $R_{31}^B$ and $R_{42}^B$ versus $R_{12}^B$ on the pseudo-critical
line, calculated from a NNLO Taylor series. Data are results on cumulant ratios of net-proton 
number fluctuations obtained by the STAR Collaboration \cite{Adam:2020unf}. Also shown
are preliminary results obtained at $\sqrt{s_{NN}} = 54.4$ GeV \cite{Pandav:2020uzx}. Dashed
lines show joint fits to the data.}
\label{fig:RnmB}
\end{figure}

As higher order cumulants are expected to diverge more rapidly when approaching a critical point, 
it is tempting to discuss also the ratios $R_{31}^B$ and $R_{42}^B$ along the pseudo critical line,
which are shown in Fig.~\ref{fig:RnmB} as a function of $R_{12}^B$ \cite{Bazavov:2020bjn}. 
Since $R_{12}^B$ is still a monotonous function of $\mu_B$ in the plotted range, it is a measure 
for the baryon density and enables us to compare with the experiment in a model free way.  
We see that the over all agreement with the corresponding net-proton number cumulants $R_{31}^P$ and
$R_{42}^P$ from STAR \cite{Adam:2020unf, Pandav:2020uzx} is very good. We conclude that a high
freeze-out temperature of $T^f>155$ MeV seems to be excluded by the data. This lattice calculation is
based on an $8^{th}$ order expansion of the logarithm of the partition function. 

Finally we want to mention that the radius of convergence, which is inherent to the 
expansion of any thermodynamic observable, can in principle provide valuable information 
on the phase structure of QCD. \textit{E.g.}, in the case of a second order phase transition,
we expect the convergence radius to be limited by the critical point. A simple estimator for the
radius of convergence $\hat\rho\equiv\mu_B^{\textrm{crit}}/T$ is given by the ratio estimator
\begin{equation}
    \hat\rho=\lim_{n\to\infty}\sqrt{(n+2)(n+1)\left|\chi_n^B/\chi_{n+2}^B\right|},
\end{equation}
more advanced estimators are also discussed \cite{Giordano:2019slo}. Unfortunately, we have only
a limited number of expansion coefficients (cumulatns $\chi_n^B$) at our disposal, which makes
it difficult to draw strong conclusions with given lattice data. Especially, since the
statistical and
systematical error on higher order cumulants is drastically increasing with the order $n$. It
is however interesting to note that all expansion coefficients have to be positive if the
limiting singularity lies on the real axis. Hence, we can obtain an
upper bound for the phase transition temperature $T_{cep}$, as for $T>140$ MeV many of the expansion
coefficients turn negative \cite{Karsch:2019mbv}. This estimate is in good agreement with the statement
that the temperature of the QCD critical point shall be lower than the chiral transition temperature
($T_{cep}<T_c$) as indicated in Fig.~\ref{fig:pdiag} (left).

\section{Cumulants at imaginary chemical potential}
\begin{figure}
    \centering
    \includegraphics[width=0.48\textwidth]{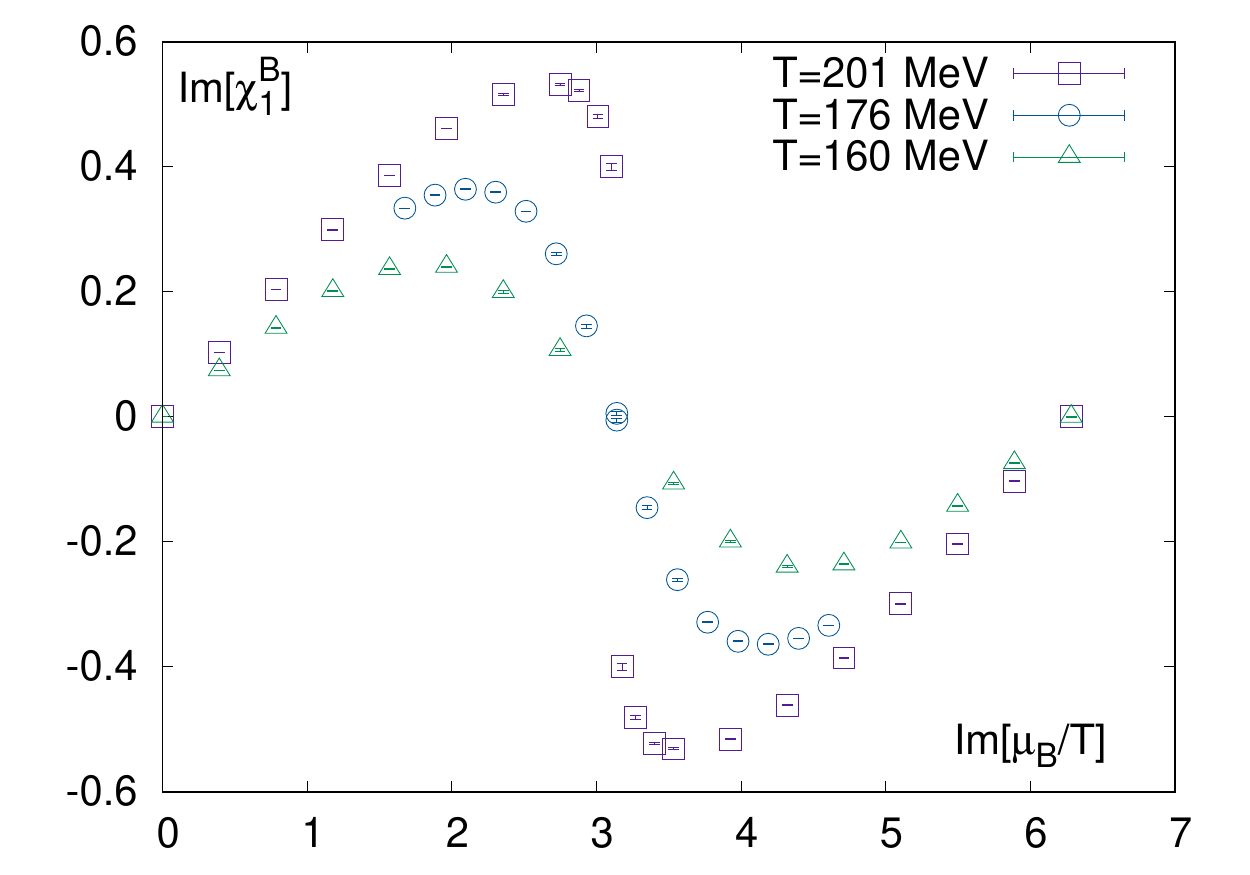}
    \includegraphics[width=0.48\textwidth]{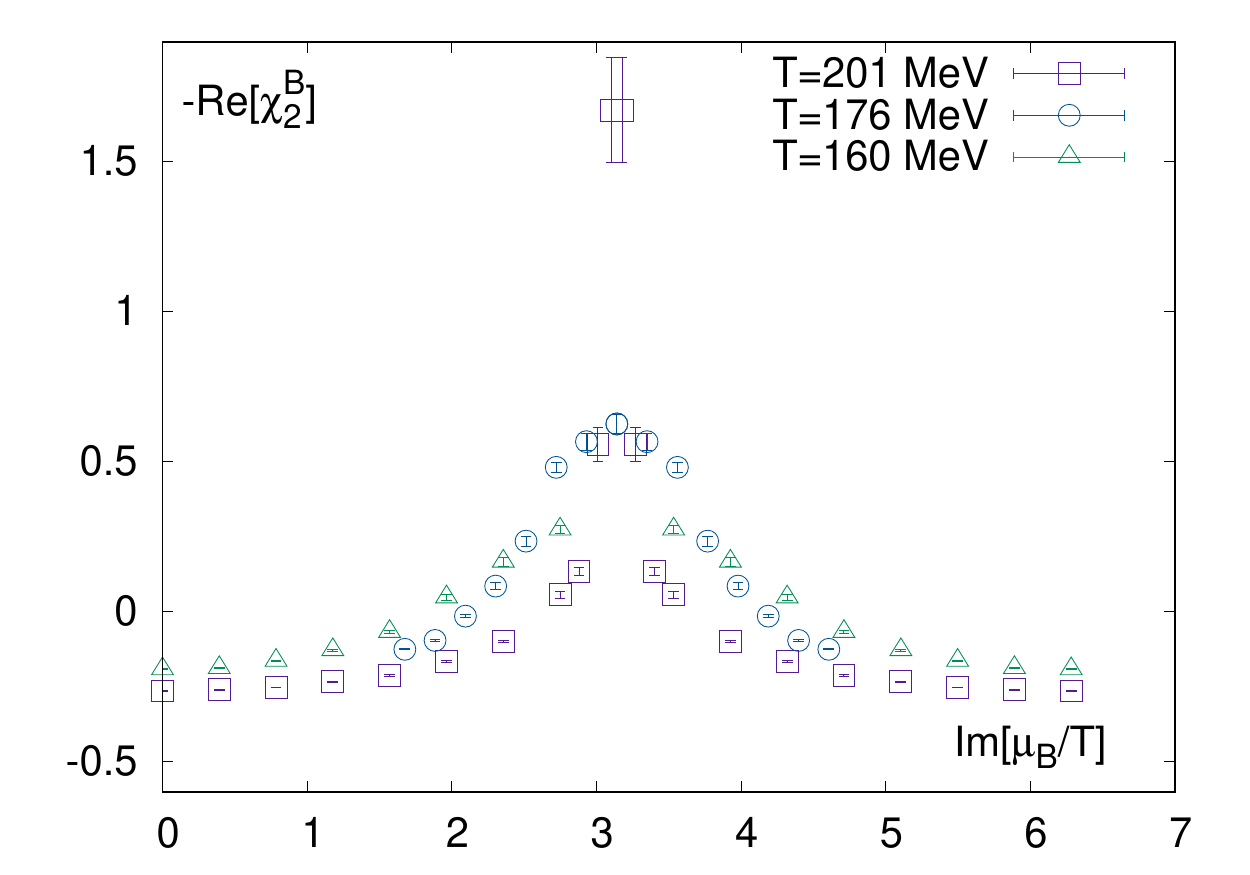}
    \caption{Preliminary results of the first and second cumulant of the net-baryon number,
    $\chi_1^B$, $\chi_2^B$, as a function of the imaginary chemical potential for three
    different temperatures, obtained from calculations on $24^3\times 4$ lattices. }
    \label{fig:imagmu}
\end{figure}

Besides the Taylor expansion method, lattice QCD calculations can also be performed at purely
imaginary
chemical potential, followed by an analytic continuation of the results. The QCD
partition function is symmetric under the transformation $\hat\mu_B\to\hat\mu_B+2\pi i$. Any
simulations at imaginary chemical potential are thus constrained to the interval 
$[-i\pi,i\pi]$ (first Roberge-Weiss sector). We further note that even/odd order cumulants on
this interval are purely real/imaginary and are even/odd functions of $\textrm{Im}[\hat\mu_B]$.
Making use of this symmetry, we thus need to simulate only in the interval $[0,i\pi]$ and
symmetrize/anti-symmetrize the data afterwards. We calculate the first four cumulants of the
baryon number. Preliminary results from $24^3\times4$ lattices are shown in
Fig.~\ref{fig:imagmu}.
We can see that the (purely imaginary) baryon number density $\textrm{Im}[\chi_1^B]$ develops a
discontinuity at $\textrm{Im}[\hat\mu_B]=\pi$. The temperature where this is happening is called the
Roberge-Weiss temperature ($T_{RW}$), which was estimated to $T_{RW}=201$ MeV \cite{Goswami:2018qhc}
(for $N_\tau=4)$. 
In accordance with this discontinuity, we also observe that the second cumulant $\chi_2^B$ develops a
divergence at $T_{RW}$. The universal scaling of the Polyakov-Loop (order parameter of the confinement
transition) and chiral condensate have been investigated close to the Roberge-Weiss transition
\cite{Goswami:2018qhc}. 

The periodic data on $\chi_1^B$ can be analyzed in terms of Fourier coefficients
\cite{Vovchenko:2017gkg, Almasi:2018lok}, which are inherently linked to the canonical
partition sums. The aim of this project is, however, to use
the information of all the available cumulants to construct a precise rational function
approximation, \textit{i.e.} a $[n,m]$ Padé of $\chi_1^B$,
\begin{equation}
\chi_1^B \approx \mathcal{R}^{m}_{n}(\hat\mu_B) = \frac{P_m}{Q_n}
= \frac{\sum_{i=0}^m \, a_i \, \hat\mu_B^i}{\sum_{j=0}^n \, b_j \, \hat\mu_B^j}.
\end{equation}
We are currently testing several methods to determine the coefficients $a_i,b_j$. Among them is
a direct solve method, where we directly solve a set of equations that we obtain by equating the
analytic expressions for $\mathcal{R}^m_n$ as well as its first few derivatives $\partial^j
\mathcal{R}^m_n /\partial \hat\mu_B^j$, $j=0,1,2$ at each simulation point with our lattice data,
\begin{align}
    \chi_{j+1}^B(\hat\mu_B^{(k)})
    =\left. \frac{\partial^j \mathcal{R}^{m}_{n}(\hat\mu_B)}{\partial
    \hat\mu_B^j}\right|_{\hat\mu_B=\hat\mu_B^{(k)}}.
\end{align}
Here $\chi_j^B(\hat\mu_B^{(k)})$ represent the numerical values of the cumulants at the simulation
points $\hat\mu_B^{(k)}$, as obtained by our lattice calculations. A similar method is based on a
$\chi^2$-fit of $\mathcal{R}^{m}_{n}$ to our cumulant data. Finally we are testing a two step approach
where in a first step a suitable interpolation of the lattice data is chosen. In the second step we are
making use of the Remez algorithm to determine $\mathcal{R}^{m}_{n}$ until the min-max criteria is
satisfied with respect to the interpolation.

Having the approximation $\mathcal{R}^{m}_{n}$ at hand, we are able to integrate the baryon density
to obtain the free energy, which will also develop a cusp at $T_{RW}$. However, our main interest
lies in the determination of the roots of the numerator $P_m$ and denominator $Q_n$, which will
allow us to infer information on the singularities in the complex $\hat\mu_B$ plane. A
singularity in the complex plane is the reason for a finite radius of convergence of the Taylor
series and will also indicate a true physical phase transition when it approaches the real axis
in the complex $\hat\mu_B$-plane. 

\begin{figure}
    \begin{center}
    \includegraphics[width=0.48\textwidth]{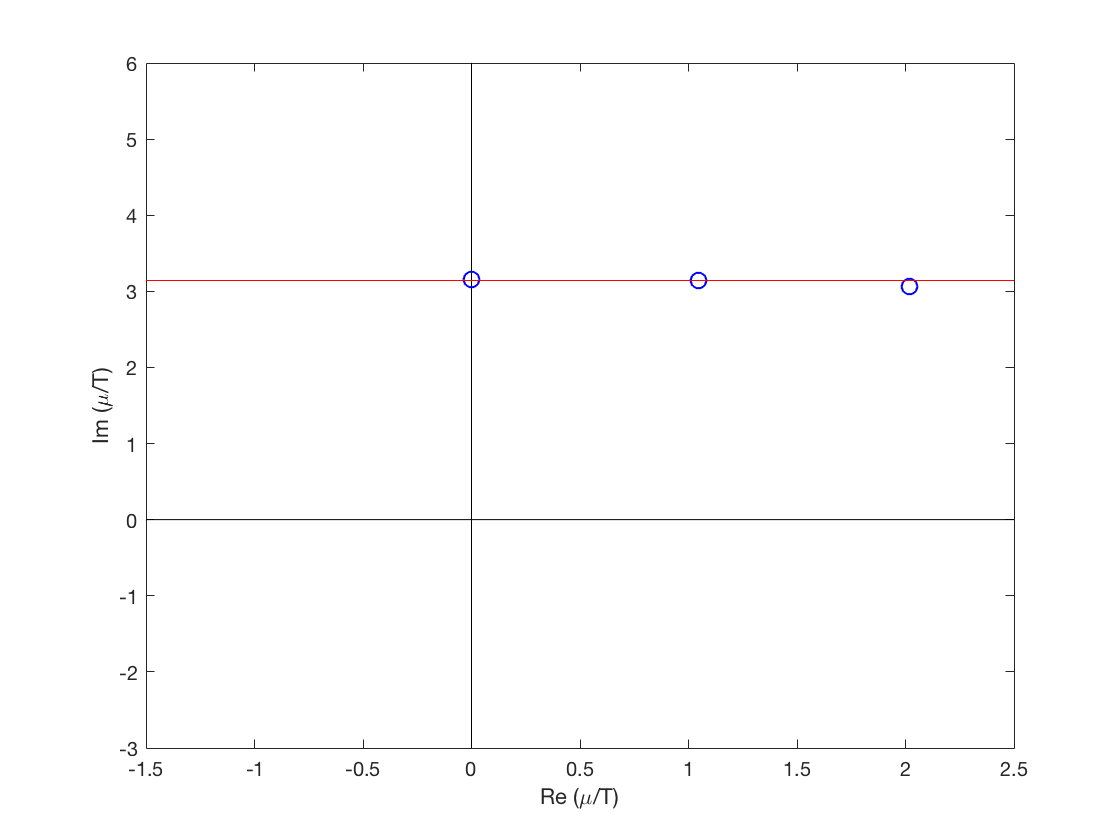}
    \includegraphics[width=0.48\textwidth]{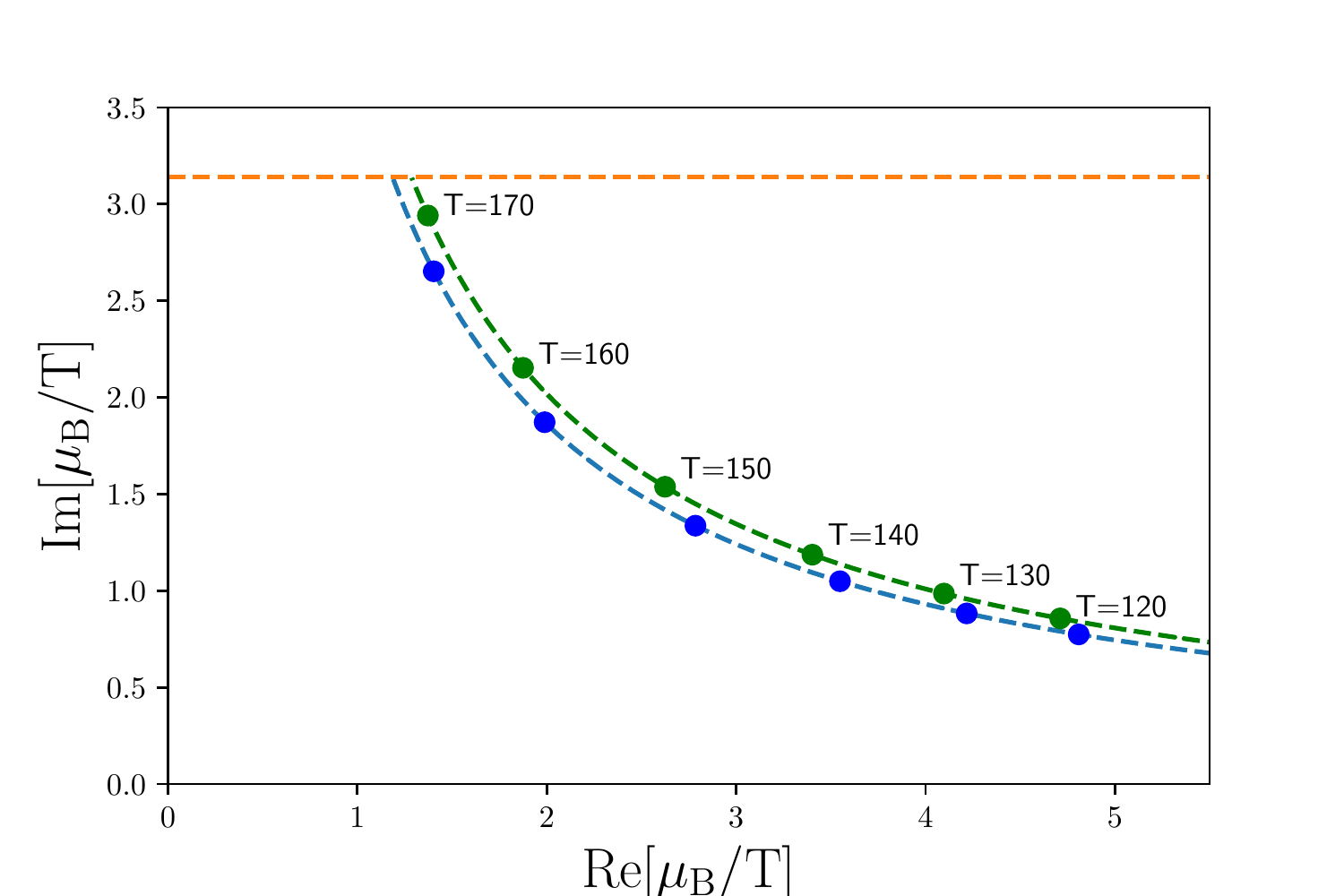}
    \end{center}
    \caption{Left: Singularity in the complex plane, associated with the branch cut singularity of the Fermi-Dirac distribution function of the quarks. The position of the singularity is shown for three temperatures $T=201, 176$ and $160$ MeV from left to right. The results have been obtained from calculations on $24^3\times 4$ lattices. Right: Singularity in the complex plane, associated to the pole in the scaling function $f_f(z)$. The results are model predictions for $N_\tau=4,6$ (blue, green) based on a mapping of QCD to the universal 3d $O(2)$-model. }
    \label{fig:poles}
\end{figure}

There are two models that can guide our thinking about the
location of the singularities in the complex plane. At large temperatures the thermal branch cut
singularities from the Fermi-Dirac distribution of a free quark gas is expected to pinch the
imaginary axis in the complex $\hat\mu_B$-plane. In QCD such a behavior is expected to happen at
$T_{RW}$. In fact, this is something we already see, when we analyze the data shown in
Fig.~\ref{fig:imagmu}. How this thermal singularity moves in the complex plane with with
decreasing temperature is shown in Fig.~\ref{fig:poles} (left).

At temperatures close to the chiral transition ($T_c$), we might be able to map our
results to the universal scaling behavior connected to the chiral phase transition. 
The scaling function of the free energy $f_f(z)$ will have a singularity in the complex
$z$-plane, known as the Lee-Yang edge singularity. This singularity has been determined
recently \cite{Connelly:2020gwa}. Given a mapping from QCD to the universal theory,
defined by the non-universal constants $t_0,h_0,T_c, \kappa_2^B$ [HotQCD, private
communication], we can calculate the position of the singularity in the complex
$\hat\mu_B$-plane, shown in Fig.~\ref{fig:poles} (right). Preliminary results from
calculations on $36^3\times 6$ lattices at $T=145$ MeV seem to be in
rather good agreement with this prediction. It will be very interesting but also
challenging to see if the singularity will approach the real axis in the complex
$\hat\mu_B$-plane for even smaller temperatures.

\section*{Acknowledgements}
We thank all members of the HotQCD collaboration for discussions and comments. 
This work is support by Deutsche Forschungsgemeinschaft (DFG,
German Research Foundation) through the Collaborative Research Centre CRC-TR 211
``Strong-interaction matter under extreme conditions'' project number 315477589 and from the
European Union’s Horizon 2020 research and innovation program under the 
Marie Skłodowska-Curie grant agreement No H2020-MSCAITN-2018-813942 (EuroPLEx). 
\printbibliography

@article{Ding:2019prx,
    author = "Ding, H.T. and others",
    title = "{Chiral Phase Transition Temperature in ( 2+1 )-Flavor QCD}",
    eprint = "1903.04801",
    archivePrefix = "arXiv",
    primaryClass = "hep-lat",
    doi = "10.1103/PhysRevLett.123.062002",
    journal = "Phys. Rev. Lett.",
    volume = "123",
    number = "6",
    pages = "062002",
    year = "2019"
}

@article{Goswami:2018qhc,
    author = "Goswami, Jishnu and Karsch, Frithjof and Lahiri, Anirban and Schmidt, Christian",
    title = "{QCD phase diagram for finite imaginary chemical potential with HISQ fermions}",
    eprint = "1811.02494",
    archivePrefix = "arXiv",
    primaryClass = "hep-lat",
    doi = "10.22323/1.334.0162",
    journal = "PoS",
    volume = "LATTICE2018",
    pages = "162",
    year = "2018"
}

@article{Bazavov:2018mes,
    author = "Bazavov, A. and others",
    collaboration = "HotQCD",
    title = "{Chiral crossover in QCD at zero and non-zero chemical potentials}",
    eprint = "1812.08235",
    archivePrefix = "arXiv",
    primaryClass = "hep-lat",
    doi = "10.1016/j.physletb.2019.05.013",
    journal = "Phys. Lett. B",
    volume = "795",
    pages = "15--21",
    year = "2019"
}

@article{Andronic:2017pug,
    author = "Andronic, Anton and Braun-Munzinger, Peter and Redlich, Krzysztof and Stachel, Johanna",
    title = "{Decoding the phase structure of QCD via particle production at high energy}",
    eprint = "1710.09425",
    archivePrefix = "arXiv",
    primaryClass = "nucl-th",
    doi = "10.1038/s41586-018-0491-6",
    journal = "Nature",
    volume = "561",
    number = "7723",
    pages = "321--330",
    year = "2018"
}

@article{Bazavov:2017dus,
    author = "Bazavov, A. and others",
    title = "{The QCD Equation of State to $\mathcal{O}(\mu_B^6)$ from Lattice QCD}",
    eprint = "1701.04325",
    archivePrefix = "arXiv",
    primaryClass = "hep-lat",
    doi = "10.1103/PhysRevD.95.054504",
    journal = "Phys. Rev. D",
    volume = "95",
    number = "5",
    pages = "054504",
    year = "2017"
}

@article{Adamczyk:2017iwn,
    author = "Adamczyk, L. and others",
    collaboration = "STAR",
    title = "{Bulk Properties of the Medium Produced in Relativistic Heavy-Ion Collisions from the Beam Energy Scan Program}",
    eprint = "1701.07065",
    archivePrefix = "arXiv",
    primaryClass = "nucl-ex",
    doi = "10.1103/PhysRevC.96.044904",
    journal = "Phys. Rev. C",
    volume = "96",
    number = "4",
    pages = "044904",
    year = "2017"
}

@article{Bazavov:2012vg,
    author = "Bazavov, A. and others",
    title = "{Freeze-out Conditions in Heavy Ion Collisions from QCD Thermodynamics}",
    eprint = "1208.1220",
    archivePrefix = "arXiv",
    primaryClass = "hep-lat",
    doi = "10.1103/PhysRevLett.109.192302",
    journal = "Phys. Rev. Lett.",
    volume = "109",
    pages = "192302",
    year = "2012"
}

@article{Adam:2020unf,
    author = "Adam, J. and others",
    collaboration = "STAR",
    title = "{Net-proton number fluctuations and the Quantum Chromodynamics critical point}",
    eprint = "2001.02852",
    archivePrefix = "arXiv",
    primaryClass = "nucl-ex",
    month = "1",
    year = "2020"
}

@article{Pandav:2020uzx,
    author = "Pandav, Ashish",
    collaboration = "STAR",
    title = "{Measurement of cumulants of conserved charge multiplicity distributions in Au +Au collisions from the STAR experiment}",
    eprint = "2003.12503",
    archivePrefix = "arXiv",
    primaryClass = "nucl-ex",
    doi = "10.1016/j.nuclphysa.2020.121936",
    journal = "Nucl. Phys. A",
    volume = "1005",
    pages = "121936",
    year = "2021"
}

@article{Bazavov:2020bjn,
    author = "Bazavov, A. and others",
    title = "{Skewness, kurtosis, and the fifth and sixth order cumulants of net baryon-number distributions from lattice QCD confront high-statistics STAR data}",
    eprint = "2001.08530",
    archivePrefix = "arXiv",
    primaryClass = "hep-lat",
    doi = "10.1103/PhysRevD.101.074502",
    journal = "Phys. Rev. D",
    volume = "101",
    number = "7",
    pages = "074502",
    year = "2020"
}

@article{Vovchenko:2017gkg,
    author = "Vovchenko, Volodymyr and Steinheimer, Jan and Philipsen, Owe and Stoecker, Horst",
    title = "{Cluster Expansion Model for QCD Baryon Number Fluctuations: No Phase Transition at $\mu_B / T < \pi$}",
    eprint = "1711.01261",
    archivePrefix = "arXiv",
    primaryClass = "hep-ph",
    doi = "10.1103/PhysRevD.97.114030",
    journal = "Phys. Rev. D",
    volume = "97",
    number = "11",
    pages = "114030",
    year = "2018"
}

@article{Almasi:2018lok,
    author = "Almasi, Gabor Andras and Friman, Bengt and Morita, Kenji and Lo, Pok Man and Redlich, Krzysztof",
    title = "{Fourier coefficients of the net-baryon number density and chiral criticality}",
    eprint = "1805.04441",
    archivePrefix = "arXiv",
    primaryClass = "hep-ph",
    doi = "10.1103/PhysRevD.100.016016",
    journal = "Phys. Rev. D",
    volume = "100",
    number = "1",
    pages = "016016",
    year = "2019"
}

@article{Connelly:2020gwa,
    author = "Connelly, Andrew and Johnson, Gregory and Rennecke, Fabian and Skokov, Vladimir",
    title = "{Universal Location of the Yang-Lee Edge Singularity in $O(N)$ Theories}",
    eprint = "2006.12541",
    archivePrefix = "arXiv",
    primaryClass = "cond-mat.stat-mech",
    doi = "10.1103/PhysRevLett.125.191602",
    journal = "Phys. Rev. Lett.",
    volume = "125",
    number = "19",
    pages = "191602",
    year = "2020"
}

@article{Giordano:2019slo,
    author = "Giordano, Matteo and P\'asztor, Attila",
    title = "{Reliable estimation of the radius of convergence in finite density QCD}",
    eprint = "1904.01974",
    archivePrefix = "arXiv",
    primaryClass = "hep-lat",
    doi = "10.1103/PhysRevD.99.114510",
    journal = "Phys. Rev. D",
    volume = "99",
    number = "11",
    pages = "114510",
    year = "2019"
}

@article{Borsanyi:2020fev,
    author = "Borsanyi, Szabolcs and Fodor, Zoltan and Guenther, Jana N. and Kara, Ruben and Katz, Sandor D. and Parotto, Paolo and Pasztor, Attila and Ratti, Claudia and Szabo, Kalman K.",
    title = "{QCD Crossover at Finite Chemical Potential from Lattice Simulations}",
    eprint = "2002.02821",
    archivePrefix = "arXiv",
    primaryClass = "hep-lat",
    doi = "10.1103/PhysRevLett.125.052001",
    journal = "Phys. Rev. Lett.",
    volume = "125",
    number = "5",
    pages = "052001",
    year = "2020"
}

@article{Attanasio:2020spv,
    author = {Attanasio, Felipe and J\"ager, Benjamin and Ziegler, Felix P.G.},
    title = "{Complex Langevin simulations and the QCD phase diagram: Recent developments}",
    eprint = "2006.00476",
    archivePrefix = "arXiv",
    primaryClass = "hep-lat",
    reportNumber = "CP3-Origins-2020-08 DNRF90, NT@UW-20-27",
    doi = "10.1140/epja/s10050-020-00256-z",
    journal = "Eur. Phys. J. A",
    volume = "56",
    number = "10",
    pages = "251",
    year = "2020"
}

@article{Alexandru:2020wrj,
    author = "Alexandru, Andrei and Basar, Gokce and Bedaque, Paulo F. and Warrington, Neill C.",
    title = "{Complex Paths Around The Sign Problem}",
    eprint = "2007.05436",
    archivePrefix = "arXiv",
    primaryClass = "hep-lat",
    month = "7",
    year = "2020"
}

@article{Karsch:2019mbv,
    author = "Karsch, Frithjof",
    editor = "Anagnostopoulos, Konstantinos and others",
    title = "{Critical behavior and net-charge fluctuations from lattice QCD}",
    eprint = "1905.03936",
    archivePrefix = "arXiv",
    primaryClass = "hep-lat",
    doi = "10.22323/1.347.0163",
    journal = "PoS",
    volume = "CORFU2018",
    pages = "163",
    year = "2019"
}
\end{document}